\documentclass[aps,pra,twocolumn,preprintnumbers,superscriptaddress]{revtex4}
\usepackage{amsmath}
\usepackage{amsfonts}
\usepackage{amsthm}
\usepackage{amssymb}
\usepackage{url}
\usepackage{hyperref}
\usepackage{ifthen}
\usepackage{color}
\usepackage{graphicx}
\newboolean{qcircuit}

\setboolean{qcircuit}{true}     %  Qcircuit

\newcommand{\ket}[1]{\left |  #1 \right \rangle}
\newcommand{\bra}[1]{\left \langle #1 \right |}
\newcommand{\Tr}{\operatorname{Tr}}
\newcommand{\comment}[1]{ }
\newcommand{\ba}{\begin{eqnarray}}
\newcommand{\ea}{\end{eqnarray}}

\newcommand{\balpha}{\boldsymbol \alpha}
\newcommand{\bbeta}{\boldsymbol \beta}

\newcommand{\ubalpha}{\boldsymbol \alpha}
\newcommand{\ubbeta}{\boldsymbol \beta}
\newcommand{\x}{\text{x}}
\newcommand{\y}{\text{y}}

\newcommand{\jdb}[1]{#1}

\begin{document}

\title{More Randomness from the Same Data}
%Device-Independent Randomness Certification from Observed Statistics

\author{Jean-Daniel Bancal}
\affiliation{Centre for Quantum Technologies, National University of Singapore, 2 Science Drive 3, Singapore 117543}
\author{Lana Sheridan}
\affiliation{Centre for Quantum Technologies, National University of Singapore, 2 Science Drive 3, Singapore 117543}
\author{Valerio Scarani}
\affiliation{Centre for Quantum Technologies, National University of Singapore, 2 Science Drive 3, Singapore 117543}
\affiliation{Department of Physics, National University of Singapore, 3 Science Drive 2, Singapore 117542}

\begin{abstract}
Correlations that cannot be reproduced with local variables certify the generation of private randomness. Usually, the violation of a Bell inequality is used to quantify the amount of randomness produced. %However, the nonlocal correlations observed in an experiment can contain more randomness than the amount certified by the violation of a specific Bell inequality. 
Here, we show how private randomness generated during a Bell test can be directly quantified from the observed correlations, without the need to process these data into an inequality. The frequency with which the different measurement settings are used during the Bell test can also be taken into account. This improved analysis turns out to be very relevant for Bell tests performed with a finite collection efficiency. In particular, applying our technique to the data of a recent experiment [Christensen \textit{et al.}, Phys. Rev. Lett. 111, 130406 (2013)], we show that about twice as much randomness as previously reported can be potentially extracted from this setup.
%about twice as much randomness per use of the setup can potentially be extracted compared to previously reported.
%one can extract about twice as much randomness as previously reported.
\end{abstract}

\maketitle

%\tableofcontents
%\newpage

%%%%%%%%%%%%%%%%%%%%%%%%%%%%%%%%%%%%%%%%%%%%%%
\section{Randomness from Bell tests}
%\label{sec:intro}

\subsection{Introduction}

Sources of true randomness have numerous applications, be they in algorithms, gambling, sampling, or cryptography~\cite{Vadhan}. However, the unpredictability of these sources is typically difficult to certify since they may be correlated to external variables, and access to those other variables could render them predictable.  Therefore, there is strong motivation for developing sources of randomness that can be certified as being uncorrelated to any outside process or variable, i.e. sources of private randomness. Quantum physics offers this opportunity when a Bell inequality is violated. Since this link between nonlocal correlations and randomness was highlighted~\cite{Colbeckthesis}, it was used for randomness generation or expansion~\cite{PAM10,PM11,FGS11,VV12} and randomness amplification~\cite{CR12,GMTDAA12,MP13}.

All previous works certify randomness through the amount of violation of one or more Bell inequalities~\cite{Pawlowski13}. As such, they only take into account a coarse grained description of the knowledge available in a device-independent experiment. Moreover, in the case of randomness generation \jdb{or expansion}, they quantified the amount of randomness by analysing each measurement setting separately (or by computing lower bounds that assume a biased choice of measurement settings). Here, we propose a method to certify rates of randomness generation in the limit of large aquisitions directly from the full expected statistics, i.e. both the outcomes correlations and the frequency with which measurement settings are chosen. We show that this tool provides immediate improvements on the quantification of private randomness by applying it in the context of Bell tests with finite detection efficiency.

\subsection{Working assumptions}
%{\bf Working assumptions.}

Randomness can be studied under different sets of assumptions. The situation we have in mind here is the generic case of an experimental physicist who wants to convince a colleague that his setup is doing its job properly. We thus choose our working assumptions accordingly.

Namely, we consider in this paper that the devices used to produce randomness were acquired from a trusted provider (or built by the trustable experimental physicist himself), meaning that the devices may not operate as hoped for, but they are precluded from being actively malicious: for instance, they don't incorporate an RF transmitter that could leak supposedly private information to the outside world. There is also no adversary Eve in this scenario, but only a skeptic verifier Thomas (the colleague), who wants unquestionable evidence that genuine private randomness is being produced. 
Thomas does not hold quantum side information about the devices (the honest provider didn't purposely design the devices so that they would remain entangled to a third system); but he may hold a more detailed classical description of the devices than that available to average users and to the provider. If this classical information is sufficient to predict the outcomes produced by the devices, because they act in a deterministic way for instance, then Thomas will conclude that there is no intrinsic randomness. In particular, pseudo-random sources would be detected in this way, since they are totally predictable given their classical description. This \textit{trusted-provider assumption} was first explicitly stated in~\cite{PM11}, and has been used to assess recent experiments on randomness expansion~\cite{PAM10,Christensen13}. 

An important concern in randomness \jdb{analysis} is \textit{measurement independence}, i.e. that measurement settings chosen in each run are independent of the state of the quantum systems measured in that run. Measurement independence can only be guaranteed with some element of trust; it has been assumed in the first studies, then relaxed in several works~\cite{H11,BG11,CR12,GMTDAA12,KHSPMKSE12,LSS13,MP13}. In our trusted-provider scenario, full measurement independence holds as a natural consequence, even when the settings are chosen with a public pseudorandom string (and can thus be known by the verifier), because one trusts that the provider didn't build the devices with any particular string of measurement choices in mind. \jdb{In this context one can thus talk of private randomness \textit{generation}, because the randomness certified here can be generated without using any initial private randomness. This contrasts with device-independent randomness expansion (DIRE) schemes, which rely on the consumption of initial private randomness~\cite{PM11}.}

%, because its generation needs not consume any private randomness.because no private randomness is necessary to generate any randomness certified in this context. This contrast with usual device-independent randomness expansion (DIRE) schemes which }

Finally, for simplicity, we also assume in this paper that all measurements are performed a large enough number of times so that the corrections of finite statistics can be safely neglected. In this asymptotic limit and under measurement independence, memory effects do not matter if the observed statistics behave like a supermartingale~\cite{Gill03}. Assuming that this will be the case in a practical realization, we can formally restrict ourselves to the case where subsequent runs of the experiment produce independent and identically distributed (i.i.d.) variables. Finite statistics corrections, even accounting for memory effects, could be implemented on top of these results as demonstrated in~\cite{PAM10,BCHK+02,Gill03}.

%%%%%%%%%%%%%%%%%%

\section{Randomness from observed statistics}
%\label{sec:randprobs}
%\ \\
%\noindent {\bf Randomness from observed correlations.} 
It is common, and convenient, to estimate the amount of private randomness that can be extracted from some correlations in terms of a Bell inequality violation~\cite{PAM10}. Indeed, violation of a Bell inequality certifies that no pre-established strategy can reproduce the observed correlations, and thus, that no one can perfectly predict them. However, a Bell inequality violation only constitutes a partial knowledge of all the statistics produced during a Bell type experiment. Since in general the full correlation statistics can be known, we ask here how much randomness can be certified based on this more complete knowledge. %This provides the largest amount of randomness that can be certified private for a given experiment under the device-independent assumptions (i.e. only from knowledge of observed statistics).

\subsection{Full observed correlations}

Just as quantum states can exhibit two kinds of randomness, the intrinsic randomness of pure states, and the randomness of mixed states~\cite{PhysRevLett.108.100402}, correlations can also exhibit these two kinds of randomness~\cite{arXiv:1305.5384}, the intrinsic randomness being uncorrelated to outside systems, and thus private. For instance, any \emph{extremal} quantum correlations $P(a,b|x,y)$, which cannot be decomposed as a convex mixture of other correlations, hold intrinsic randomness: the maximum probability with which an external observer is able to guess the outcomes observed by the parties when they perform measurements $(x,y)$ is indeed given by the probabilities $P(ab|xy)$ themselves through:
\begin{equation}
G_{x,y} = \underset{a,b}{\max}\ P(ab|xy)\,.   \qquad  [P \text{ extremal}]
\end{equation}
The associated randomness is then $H_{\min}(AB|XY)=-\log_2(G_{xy})$. Here $a, b$ denote outcomes that two parties, Alice and Bob can observe when they perform measurements $x, y$ on their respective systems.

Conversely, when some correlations $P(ab|xy)$ can be decomposed as a convex mixture of other quantum correlations, one cannot a priori exclude that part of the outcomes' indeterminacy is simply due to classical lack of knowledge of Alice and Bob. In the trusted-provider assumptions mentioned earlier, which ensures that the provider of the boxes used by Alice and Bob keeps no state entangled with the devices in his factory, the quantum state shared by the parties and the verifier Thomas takes the form of a quantum-classical (q-c) state:
\begin{equation}\label{eq:state}
\rho_{ABT} = \sum_{\lambda} q_\lambda \rho_{AB}^{(\lambda)} \otimes \ket{\lambda}\bra{\lambda} \, .
\end{equation}
Here $\lambda$ is a classical variable potentially known to the verifier which can help him refine his guessing strategy. It provides the most detailed information when the $\rho_{AB}^{(\lambda)}$ are pure states. In particular, the verifier does not hold a purification of $\rho_{AB}$, all his side-information is classical. Moreover, as discussed above, $\rho_{ABT}$ can be assumed to be identical for each run.

Given the form of the state $\rho_{ABT}$, the correlation observed by the parties can be written as 
\begin{equation}\label{eq:decomposition}
P(ab|xy)=\sum_\lambda q_\lambda P_\lambda(ab|xy)\,.
\end{equation}
For an arbitrary set of correlations (i.e. possibly not extremal ones) the average probability with which Thomas can guess the outcomes observed by Alice and Bob correctly, given his knowledge of the inputs $x$, $y$ and of $\lambda$, is thus given by the maximum of
\begin{equation}\label{eq:guessingProb}
G_{x,y}(P|\{q_\lambda,P_\lambda\}_\lambda) = \sum_\lambda q_\lambda\ \underset{a,b}{\max}\ P_\lambda(ab|xy)
\end{equation}
over all convex decompositions~\eqref{eq:decomposition} that reproduce, on average, the observed correlations $P(ab|xy)$.

At first, it is not clear whether such an expression could be optimized easily, because it might involve an arbitrary number of $\lambda$'s. The following proposition is useful in this respect (proof in Appendix~\ref{proof}).

\paragraph*{Proposition 1.} Without loss of generality, the maximum value of
\begin{equation}\label{eq:proposition}
\sum_\lambda q_\lambda\ \underset{a,b}{\max} \sum_{x,y,\mu} f(x,y,\mu) P_\lambda(a(x,y,\mu),b(x,y,\mu)|xy)
\end{equation}
over all decompositions of the form~\eqref{eq:decomposition} can be achieved by considering one term in this decomposition per possible argument of the inner maximization. Here $\mu$ is a discrete variable, and $f(x,y,\mu)$, $a(x,y,\mu)$, $b(x,y,\mu)$ are functions defined on the support of $f(x,y,\mu)$.

To clarify, the arguments of the maximization are each value that the outcomes can take, given each value the inputs $x,y$, and perhaps some other variable $\mu$, can take.  So, in the specific case of evaluating equation~\eqref{eq:guessingProb}, the support of $f(x,y,\mu)$ consists of one point, and the summation over $x,y,$ and $\mu$ is effectively already subsumed into $P_\lambda$. Then if the number of possible outcomes for Alice's measurements is given by $|a|$, and similarly $|b|$ denotes Bob's number of possible outcomes, this proposition ensures that no more than $|a|\cdot|b|$ terms need to be considered in the decomposition~\eqref{eq:decomposition} when optimizing~\eqref{eq:guessingProb}. Setting $\lambda=(\alpha,\beta)$, with $\alpha=\{0\ldots|a|-1\}$, $\beta=\{0\ldots|b|-1\}$ one can thus use the following program to compute the amount of private randomness that can be extracted from the correlations $P(ab|xy)$ when settings $x,y$ are used:
\begin{equation} \label{eq:program}
\begin{split}
G_{x,y}(P) = \underset{P_{\alpha\beta}}{\ \ \max\ \ } & \sum_{\alpha\beta} P_{\alpha\beta}(\alpha\beta|xy)\\
\text{s.t.\ \ } &  \sum_{\alpha\beta} P_{\alpha\beta}(ab|xy) = P(ab|xy)\\
%& \sum_{ab\alpha\beta} P_{\alpha\beta}(ab|xy) = 1\\
& P_{\alpha\beta}(ab|xy)\text{ is quantum},
\end{split}
\end{equation}
where the sum over $\alpha$, $\beta$ indicates that there is one value of  $\lambda$ for each value of $\alpha$ and $\beta$, and we have absorbed the weights $q_{\alpha\beta}$ into the normalization of the probability distributions $P_{\alpha\beta}$.

Notice that this expression uses the observed distribution $P(ab|xy)$ directly rather than the value of a Bell inequality.

%%%%%%%%%%%%%%%%%%

\subsection{Taking the input distribution into account}
\label{sec:inputdistribution}
%\ \\
%\noindent{\bf Taking the input distribution into account.} 
So far we considered the randomness that can be certified in presence of correlations $P(ab|xy)$ when a given set of settings $(x,y)$ are used. This analysis can be straightforwardly generalized to the situation in which the inputs are chosen with some arbitrary probability $p(x,y)$ by noting that the average guessing probability in this case reads
\begin{equation}\label{eq:G}
\begin{split}
G(P|\{q_\lambda,P_\lambda\}_\lambda) &= \sum_{xy} p(x,y) G_{x,y}(P|\{q_\lambda,P_\lambda\}_\lambda)\\
&= \sum_\lambda q_\lambda \underset{{\bf a}, {\bf b} }{\max} \sum_{xy}p(x,y)P_\lambda({\bf a}_{xy} {\bf b}_{xy}|x,y)
\end{split}
\end{equation}
where ${\bf a}_{xy}\in\{0,\ldots,|a|-1\}$ and ${\bf b}_{xy}\in\{0,\ldots,|b|-1\}$ for all $x=\{0,\ldots,|x|-1\}$ and $y=\{0,\ldots,|y|-1\}$ denote the outcomes considered for each of the settings $x$, $y$.

Thanks to proposition 1 it is sufficient to consider only at most $(|a||b|)^{|x||y|}$ $\lambda$'s when optimizing \eqref{eq:G} over all decompositions of the form \eqref{eq:decomposition} (since the support of $f(x,y,\mu)=p(xy)$ in proposition 1 is $|x||y|$ distinct points).
The amount of private randomness that can be certified when some correlations are observed by choosing settings according to the distribution $p(x,y)$ can thus be quantified by the following program:

\begin{equation} \label{eq:bigProgram}
\begin{split}
G(p,P) = \underset{P_{\ubalpha\ubbeta}}{\ \ \max\ \ } & \sum_{xy}p(x,y)\sum_{\ubalpha\ubbeta} P_{\ubalpha\ubbeta}(\balpha_{xy}\bbeta_{xy}|xy)\\
\text{s.t.\ \ } &  \sum_{\ubalpha\ubbeta} P_{\ubalpha\ubbeta}(ab|xy) = P(ab|xy)\\
%& \sum_{ab\alpha\beta} P_{\alpha_{xy}\beta_{xy}}(ab|xy) = 1\\
& P_{\ubalpha\ubbeta}(ab|xy)\text{ is quantum},
\end{split}
\end{equation}
where the weights are again absorbed into the normalizations of $P_{\ubalpha \ubbeta}$ and we have $\ubalpha = \{\balpha_{00}, \ldots \balpha_{|x|-1,|y|-1}\}$ with ${\balpha}_{xy}\in\{0,\ldots,|a|-1\}$ and likewise for $\ubbeta$ and ${\bbeta}_{xy}$, i.e. bold greek letters $\balpha$ and $\bbeta$ denote vectors taking one of $|a|^{|x||y|}$ or $|b|^{|x||y|}$ possible values (as indexed by the components $\balpha_{xy}$ and $\bbeta_{xy})$.

%The idea behind this program is that Thomas could happen to hold a description of the decomposition best suited to for this case. However he does not have active access to the boxes: under measurement independence, the decomposition cannot depend on the particular measurement that is performed in each run, and so the decomposition~\eqref{eq:decomposition} must be identical for all choices of settings $x$, $y$.

%The idea behind this program is that Thomas could know \jdb{the inputs to be used by the parties. However, he does not have active access to the boxes: under measurement independence, the decomposition cannot depend on the particular measurement that is performed in each run. So the decomposition~\eqref{eq:decomposition} must be identical for all choices of settings $x$, $y$, and it can at best be adapted to the relative frequencies with which inputs are chosen.}

The idea behind this program is that Thomas could know the relative frequencies of the inputs and happen to hold a description of the best decomposition for this case. However he does not have active access to the boxes: under measurement independence, the decomposition cannot depend on the particular measurement that is performed in each run, and so the decomposition~\eqref{eq:decomposition} must be identical for all choices of settings $x$, $y$.

%%%%%%%%%%%%%%%%%%

\subsection{Getting practical bounds and certificate}
%\label{sec:randomfromprogram}

%\ \\
%\noindent {\bf Getting practical bounds and certificate.}

In order to perform the optimizations~(\ref{eq:bigProgram}) (of which (\ref{eq:program}) is a special case) one needs to describe accurately the set of quantum correlations. Since no method is known for this, we upperbound the guessing probability $G$ by relaxing the constraint that $P_{\balpha\bbeta}(ab|xy)$ be quantum, to let it belong to some level of the NPA hierarchy~\cite{NPA07,NPA08}. This turns equations~(\ref{eq:program}) and~(\ref{eq:bigProgram}) into semidefinite programs which can be efficiently computed numerically, with the guarantee of approaching the exact quantum value as the level of hierarchy increases.

%This relaxation allows to obtain the curves shown in figure \ref{fig:chsh}. These curves were moreover checked to be optimal, and are thus solution of the initial programs~\eqref{eq:program} and~\eqref{eq:bigProgram}.

Alternatively, one could also replace this condition by requiring only that the correlations $P_{\ubalpha\ubbeta}(ab|xy)$ be non-signalling. In this case the optimization takes the form of a single linear program, and describes the randomness exatractable from correlations in a world described by a generalized no-signalling theory rather than quantum theory.

Apart from computing the amount of randomness that can be extracted with respect to a verifier in different contexts, one can show that these programs also provide, through their dual, the description of a certificate for this conclusion. This certificate takes the form of a Bell expression, whose expectation value on the tested correlations $P(ab|xy)$ can only be observed in presence of the amount of randomness found by the program itself. To see this (full details can be found in Appendix~\ref{sec:dual}), we note that the optimization \eqref{eq:bigProgram} can be straightforwardly adapted to estimate the amount of randomness that can be certified from a Bell inequality violation. Namely, if $I=\sum_{abxy} c_{abxy} P(ab|xy)$ is a Bell expression then the maximum guessing probability achievable by a verifier when the Bell value $v$ is observed is given by:
\begin{equation} \label{eq:program2}
\begin{split}
G(v) = \underset{P_{\ubalpha\ubbeta}}{\ \ \max\ \ } & \sum_{xy}p(x,y)\sum_{\ubalpha\ubbeta} P_{\ubalpha\ubbeta}(\balpha_{xy}\bbeta_{xy}|xy)\\
\text{s.t.\ \ } & \sum_{abxy\ubalpha\ubbeta}c_{abxy}P_{\ubalpha\ubbeta}(ab|xy) = v\\
& \sum_{ab\ubalpha\ubbeta} P_{\ubalpha\ubbeta}(ab|xy) = 1\\
& P_{\ubalpha\ubbeta}(ab|xy)\text{ is quantum}.
\end{split}
\end{equation}

Note that a similar program using only one value of $\balpha$ and $\bbeta$ was already presented in~\cite{PAM10} to evaluate the randomness associated to a Bell inequality violation with a biased choice of inputs. However, its application relies on the concavity of the obtained function (or on its concave hull)~\cite{PhysRevLett.108.100402}, which is not always guaranteed (see Appendix~\ref{sec:concavity} for an example of non-concave function obtained in this way). The application of that program thus requires in general knowledge of the function for all violation $v$. Here, program \eqref{eq:program2} takes care of the concavity requirements through the use of several $\lambda$'s. It thus directly produces the guessing probability relevant for randomness estimation.

%Having provided new tools to estimate the randomness generated during a Bell-like experiment, we now use them to derive a Bell inequality better suited than CHSH for randomness certification in presence of finite-efficiency detectors.

\section{Results}

\subsection{Role of the inputs distribution}
\begin{figure}
\includegraphics[width=0.5\textwidth]{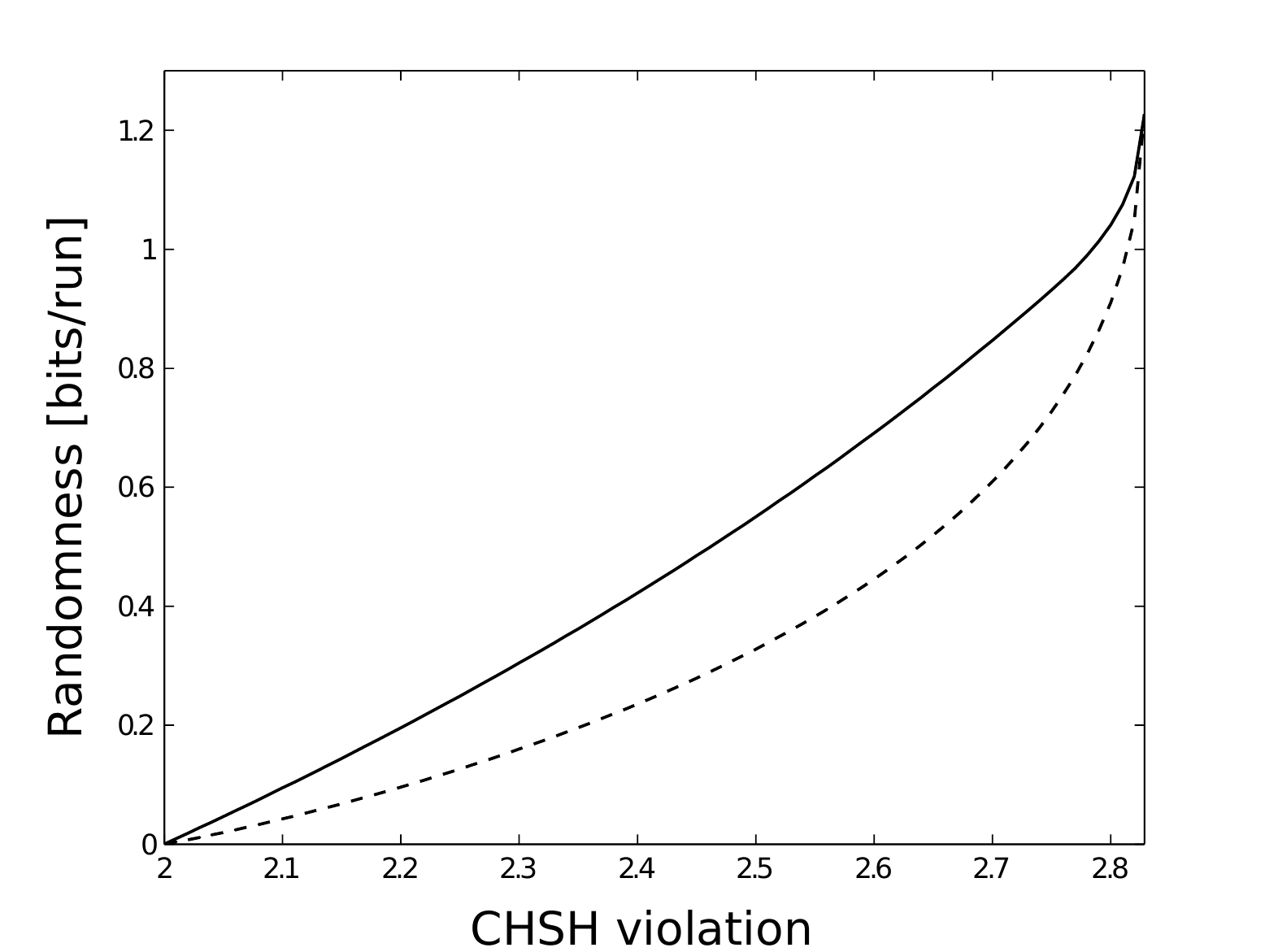}
\caption{Private randomness that can be certified as a function of CHSH violation in the fully biased input case (dashed line) and in the uniform choice of inputs case (full line).}
\label{fig:chsh}
\end{figure}

%\noindent{\bf Role of the inputs distribution.}
To illustrate the role of the inputs distribution in randomness certification, we compare in figure \ref{fig:chsh} the randomness that can be certified in presence of a CHSH violation through~\eqref{eq:program2} when one pair of settings is preferentially chosen, or when all settings are chosen with equal probability~\footnote{These curves were actually computed using an SDP relaxation as described previously, but they were checked to be optimal.}. Significantly more randomness (about twice as much) can be certified in presence of nonmaximal CHSH violation if the settings are chosen uniformly.

This can be understood by considering that when the randomness is extracted from a fixed choice of settings, one cannot exclude that the decomposition \eqref{eq:decomposition} is most adapted for the verifier to guess the outcomes observed when those settings are precisely used. However such a decomposition need not be optimal to guess the outcomes observed when performing other measurements. Since the decomposition is the same independently of which settings are used, it follows that the average guessing probability over the different settings is reduced.

%%%%%%%%%%%%%%%%%%

%\section{Application to finite efficiency Bell tests and a new Inequality}
%\label{sec:belltest}

\subsection{Application to finite efficiency Bell tests}

In practice, certifying private randomness generation in a Bell experiment requires a reasonable separation between the measured subsystems, as well as closure of the detection loophole. Without these conditions fulfilled, the possibility remains for the verifier to find a way of guessing the outcomes observed during the experiment, which then cannot be guaranteed private~\cite{PhysRevLett.107.170404}.

\begin{figure}
\includegraphics[width=0.45\textwidth]{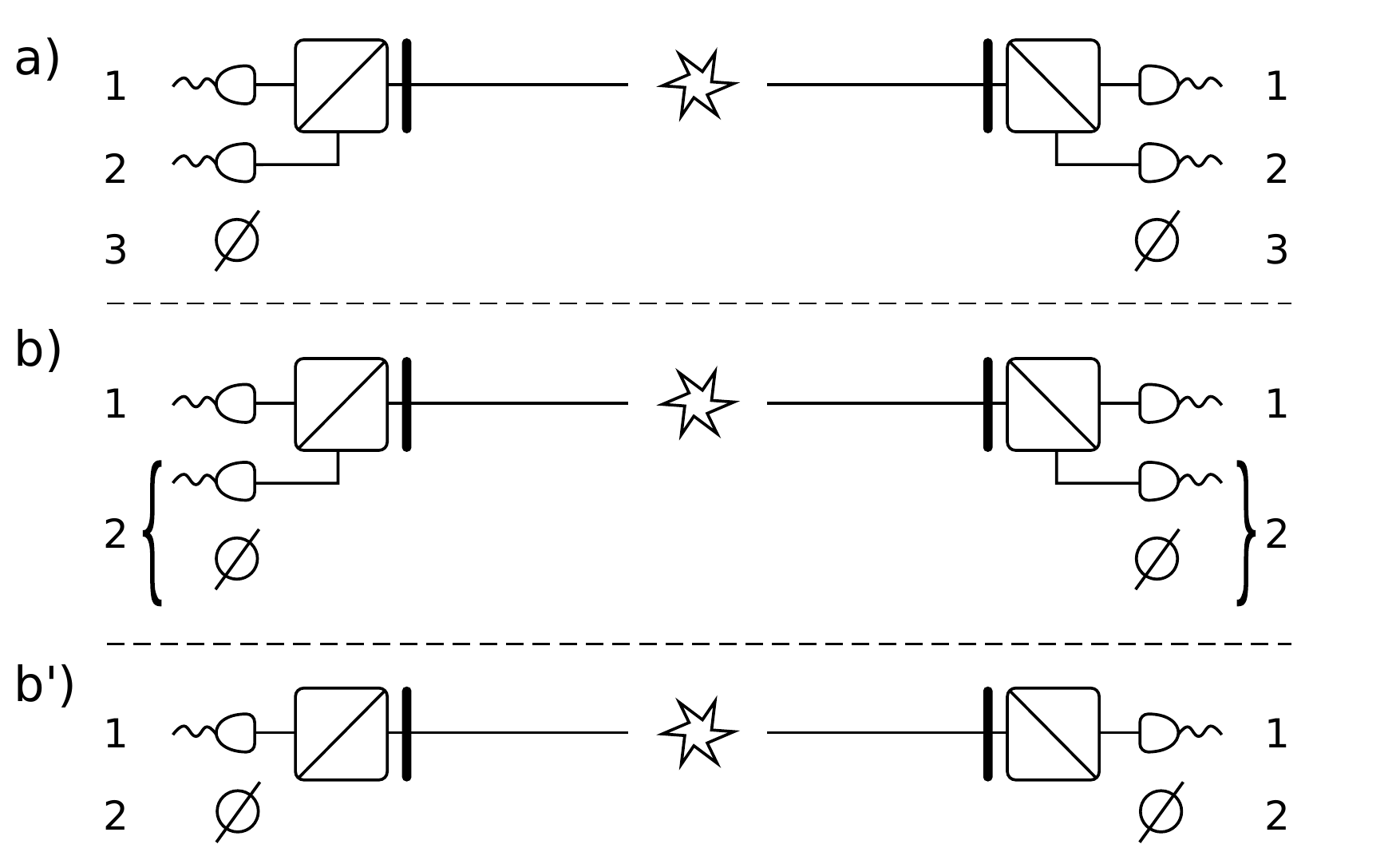}
\caption{Lossy Bell tests with different numbers of detectors and possible outcomes. a) With two detectors on each side, it is possible to attribute a distinct outcome to the result of each successful measurement, while identifying no-detection events. This is the most general treatment of lossy Bell test, but it requires the usage of three outcomes on each side. b) In order to recover two outcomes only, the no-detection events must be associated with one of the detector. b') The same statistics as produced in case b) can be obtained by using only one detector on each side. In this case, a click triggers the first outcome while the second outcome is produced in case of no-detection.}
\label{fig:setups}
\end{figure}

These constraints on parties separation and on their detection efficiency are still very demanding experimentally today. With photonic systems for instance, overall detection efficiencies which are sufficient to close the detection loophole have just been recently demonstrated~\cite{Giustina13,Christensen13}. It is thus experimentally relevant to ask what is the best way to certify private randomness in presence of finite efficiency detectors.

In the simplest Bell experiment in which two parties can each use two binary measurements, finite detection efficiency can be dealt with in several ways (see Fig.~\ref{fig:setups}). Either a third outcome is introduced for both Alice and Bob and all photons that go undetected are assigned to the third outcome, or all undetected photons are assigned to one of the two existing outcomes. The first solution is more general, but requires the usage of three-outcome Bell inequalities and two efficient detectors each for Alice and Bob, while the second solution can be treated in the two-outcome paradigm and has the advantage of only requiring one detector on each side. % (results from the other detector are assigned the same result whether detected or not)
Here we consider the two cases, with all detectors having the same efficiency $\eta=\eta_A=\eta_B$.

\begin{figure}
\includegraphics[width=0.45\textwidth]{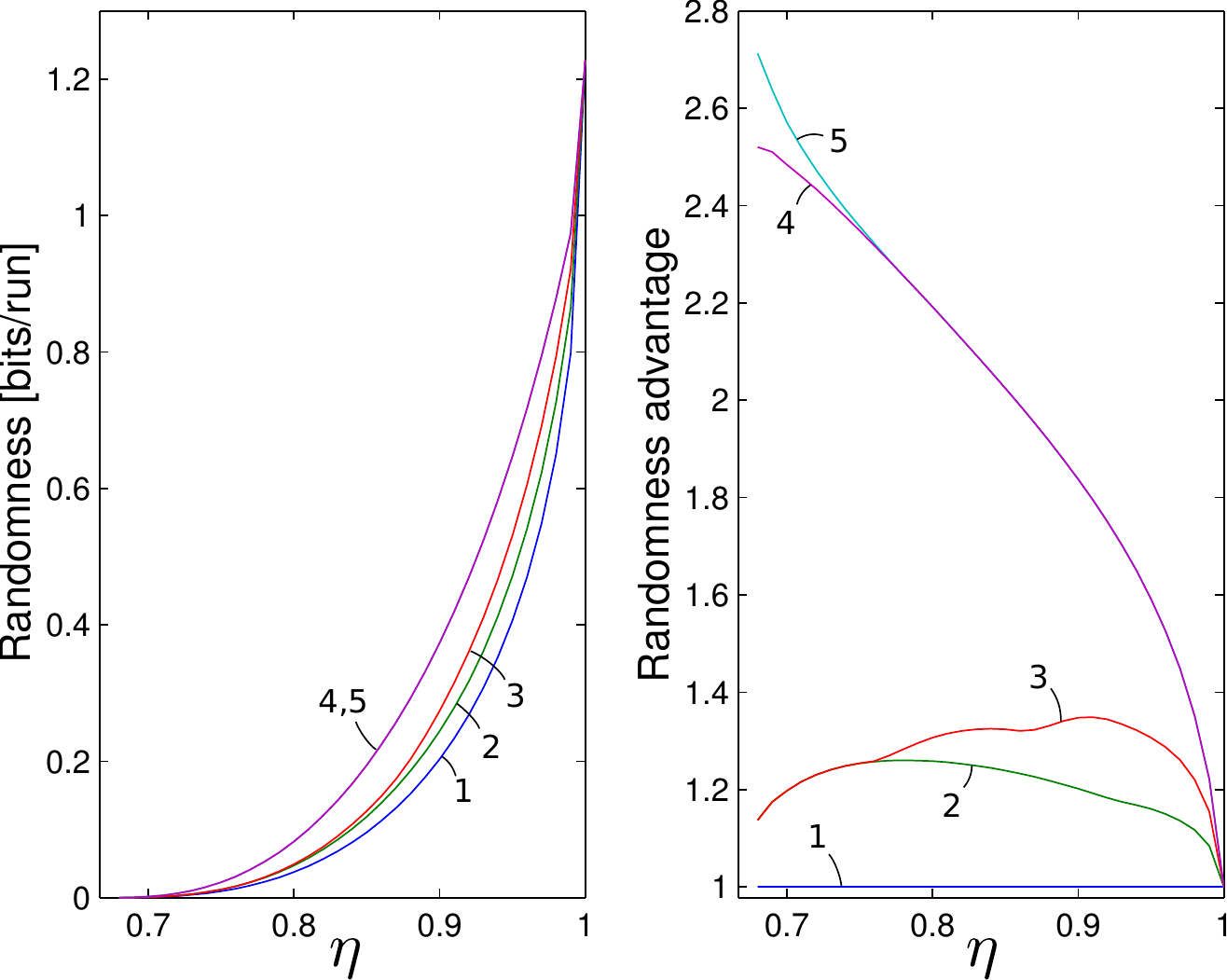}
\caption{Comparison of the randomness that can be certified in an Eberhard experiment under different circumstances. Curves 1,2,3 show the bound obtained when extracting randomness from a specific choice of settings, whereas curves 4 and 5 assume a uniform choice of settings. Curves 1 and 4 based on the observed CHSH violation were computed with~\eqref{eq:program2} (curve 1 serves as a reference with a normalization of 1 in the second figure). Curves 2 and 5 take into account the full 2-outcomes statistics and were computed with~\eqref{eq:program} and~\eqref{eq:bigProgram}. For $\eta \lesssim 90\%$ the amount of randomness displayed by curve 2 can be certified with inequality~\eqref{eq:newineq}. Curve 3, based on the full 3-outcomes statistics was computed with~\eqref{eq:bigProgram}. Bounds were computed with the Sedumi solver~\cite{S98guide} at local level 1 of the NPA hierarchy for uniform choice of settings, and with a higher level for based inputs~\cite{NPA07,Moroder13}.}
%Comparison of the randomness that can be certified in presence of finite detection efficiency by the method~\eqref{eq:program} and the Bell inequality~\eqref{eq:newineq}. 
%The blue curve plots the randomness from CHSH and biased choice of inputs.   In the second figure, this serves as the reference curve with a normalization of 1.  The green curve is the randomness from correlations and biased choice of inputs.  This is also the curve representing the randomness extractable from the inequality~\eqref{eq:newineq} with biased inputs.  The randomness from 3-outcome correlations and biased choice of inputs is plotted in red.  The purple and light blue curves show the randomness  for uniform choice of inputs, from CHSH (purple) and from correlations (light blue).  Curves are stopped where the precision of the SDP solution given by the solver (Sedumi)~\cite{S98guide} is juged insufficient.  \ls{Please check what I've claimed about the green curve.} These bounds were computed at local level 1 of the NPA hierarchy~\cite{NPA07, Moroder13}}
\label{fig:comparison}
\end{figure}

Eberhard~\cite{E93} showed that whenever $\eta>2/3$, it is possible to violate the CHSH inequality by measuring partially-entangled states of the form $\ket{\psi}=\cos\theta\ket{00}+\sin\theta\ket{11}$ with settings of the form $A_0 = \cos\alpha_1\sigma_z - \sin\alpha_1\sigma_x$, $A_1 = \cos\alpha_2\sigma_z + \sin\alpha_2\sigma_x$, $B_0 = \cos\alpha_1\sigma_z + \sin\alpha_1\sigma_x$, $B_1 = \cos\alpha_2\sigma_z - \sin\alpha_2\sigma_x$ where $\theta$, $\alpha_1$, $\alpha_2$ are functions of $\eta$. Here we analyse how much randomness can be extracted in presence of these correlations. The results are summarized in figure~\ref{fig:comparison}.

As shown in the figure, different amounts of randomness can be certified depending of the approach used. The optimal bounds are obtained when choosing measurement settings uniformly. In this case the CHSH inequality is the optimal randomness witness whenever the efficiency is $\eta \gtrsim 80\%$. 

%To demonstrate the applicability of our method to practical experiments 
To demonstrate the improvement provided by our method in presence of finite detection efficiency, we applied it to the correlations reported in~\cite{Christensen13}. The program~\eqref{eq:bigProgram} certifies a rate of 0.00014567 bits per run, \emph{i.e.} two times more than mentioned in that paper. Notice that, because of statistical fluctuations, the frequencies computed from the raw data are non-quantum and even signaling; this artefact can be corrected with a minor reformatting. (see appendix~\ref{sec:exptAnalysis} for details).

\subsection{A new Bell expression.}

Let us now go one step back and fix a highly biased inputs distribution (i.e. let's consider extracting randomness from a fixed set of inputs). When $\eta < 1$, the CHSH inequality does not certify the largest amount of randomness. The most randomness is obtained by considering the three-outcomes correlations, but already the bounds computed from the two-outcomes correlations show an improvement.

As it turns out, for efficiencies smaller than $\sim 90\%$, we can provide a Bell inequality that is optimal, in the sense that its amount of violation certifies the optimal amount of randomness. This inequality is found using the dual of~\eqref{eq:program} to be
\begin{eqnarray}
V &=& \left| \gamma \langle A_0 B_0 \rangle + \langle A_0 B_1 \rangle + \langle A_1 B_0 \rangle - \langle A_1 B_1 \rangle \right|   \nonumber \\
&\leq& \max(1+ \gamma,3-\gamma)
\label{eq:newineq}
\end{eqnarray}
where the right hand side of the expression is the bound for local correlations (computed independently of the SDP dual). Note that this inequality is not a facet of the local polytope, since this scenario only has one nontrivial family of facets, given by the CHSH inequalities~\cite{faacets.com}. Yet, it is better adapted for randomness extraction in this situation than the CHSH inequality.

%Indeed, comparing the randomness that can be certified from violation of this inequality to the one certified by violation of the CHSH inequality (which corresponds to $\gamma=1$) we see that it is larger for all $2/3< \eta<1$, and almost always optimal since it coincides with the result of~\eqref{eq:program}. Moreover, numerical optimization indicates, for detection efficiencies close to $2/3$, that no more randomness than the amount certified through~\eqref{eq:newineq} can be certified even when measurement settings are optimized (see figure~\ref{fig:comparison}).

In appendix~\ref{sec:gammaineq} we give a brief description of this inequality, and in particular give an analytical estimate of the intrinsic randomness that can be certified on the marginal correlations for this inequality.  This analysis reveals that available analytical techniques useful to certify randomness in the marginal outcomes~\cite{PAM10}, are not adapted to this inequality. The reason for this is that they treat all inputs identically, but a symmetry between the inputs is broken in inequality~\eqref{eq:newineq} whenever $\gamma\neq1$. In particular, they assume that measurements can be chosen along $\sigma_z$, whereas maximal violation of this inequality with non-maximally entangled state of the form $\cos \theta \ket{00}+\sin \theta\ket{11}$ can require measurement settings not aligned with the $z$ axis of the Bloch sphere.

%Finally, we note that using the three-outcome strategy to deal with no-detections allows to preserve more randomness (red curve in figure~\ref{fig:comparison}) in the correlations for $\eta \gtrsim 80\%$. We do not describe here the three-outcome Bell inequalities that can certify this advantage, but they can be found from the dual of~\eqref{eq:bigProgram} as mentioned previously.

%%%%%%%%%%%%%%%%%%

\section{Conclusions}
\label{sec:conclusions}

%\noindent {\bf Conclusion.}
Private randomness is a valuable resource. We have presented here a way to bound the private randomness generated during a Bell test by directly using knowledge of the full outcomes correlations and of the inputs' choice distribution.
%that can be extracted from correlations directly from the 
% \jdb{a tool that ...}
%a way to bound the private randomness that can be extracted from the correlations observed in a Bell experiment directly from the statistics which are accessible in the experiment.
Our analysis is concerned with large data sets and makes some arguably reasonable assumptions about the devices used in the Bell experiment, but does not require a characterization of these devices.  Working directly from the observed correlations allows us to find tighter bounds on the min-entropy of the output than previously known, thus certifying more randomness from identical experiments.  We demonstrate this explicitly for the results of~\cite{Christensen13}, effectively doubling the expected private randomness production rate.

This approach also furnishes new Bell inequalities through semi-definite duality to bound the guessing probability.  These inequalities may not be facets of the local polytope (inequality~\eqref{eq:newineq} is not), but they are the ones suited to randomness extraction for the form of the correlations supplied.

Even though we focused here on bipartite Bell experiments, our result generalizes directly to scenarios involving arbitrary alphabet size of inputs and outputs and any number of parties.  %This technique works either for assuming the correctness of quantum mechanics or for certifying randomness assuming only that signals do not travel faster than light.
The tools presented here can also be applied under different sets of assumptions. For instance, they can provide bounds in presence of an untrusted provider keeping a purification of the boxes if the inputs are made public only after the purification decohered.

%Note that because of the assumption we make that the source of states in the Bell test is not correlated to the inputs, the choice of the inputs need not be made with private randomness: the only important detail is the relative frequencies of the different inputs.  From this point of view, what occurs in the Bell experiment is \emph{private randomness generation}, with the certification coming from the SDP providing proof of a bound on how correlated the outputs can be with the inputs or any other external variable, given either quantum states or non-signalling ones.

%Since our bounds on the min-entropy of the output bits are less than 1 for most of the considered efficiencies, the bits produced are somewhat predicable.  Under our assumptions we can amplify the randomness of these bits using a publicly chosen classical privacy amplification hash function, since the witnessed correlations certify that each output bit cannot be very correlated with any variable held by the adversary or any other one of the output bits.  A suitable hash function will reduce the length of the input bit string, but decrease the adversary's guessing probability of each bit.

It would be relevant to extend the bounds presented here to the case of finite statistics, including devices with memory, possibly using methods similar to~\cite{PAM10,BCHK+02,Gill03}. This work also shows that new ways of obtaining analytical bounds on the randomness need to be developped to certify optimal rates of randomness generation analytically.

%%%%%%%%%%%%%%%%%%
\ \\
\noindent {\bf Note added.}
While writing this article we became aware of a similar work by O. Nieto-Silleras, S. Pironio and J. Silman~\cite{Silleras13}.

\section*{Acknowledgements}

We thank Alessandro Cere, Yun Zhi Law, Charles Ci Wen Lim and Stefano Pironio for useful discussion and comments on the manuscript. This work is funded by the Singapore Ministry of Education (partly through the Academic Research Fund Tier 3 MOE2012-T3-1-009) and the Singapore National Research Foundation.

%\bibliographystyle{unsrt}	
%\bibliography{randomness3}

\appendix

\begin{widetext}

%%%%%%%%%%%%%%%%%%

\section{Proof of propositions 1}
\label{proof}
Let us show that whenever the best choice of the function $a(x,y,\mu)$ and $b(x,y,\mu)$ for the inner optimization of \eqref{eq:proposition} is identical for two strategies $\lambda'$ and $\lambda''$, i.e.
\begin{equation}
\begin{split}
&\underset{a,b}{\text{argmax}}\ \sum_{x,y,\mu}\ f(x,y,\mu)P_{\lambda'}(a(x,y,\mu),b(x,y,\mu)|xy)\\
&=\,\underset{a,b}{\text{argmax}}\ \sum_{x,y,\mu}\ f(x,y,\mu)P_{\lambda''}(a(x,y,\mu),b(x,y,\mu)|xy),
\end{split}
\end{equation}
then the two stategies can be grouped together. This implies that it is sufficient to consider one $\lambda$ for each possible argument of this maximum.

%, and thus the number of strategies that needs to be considered in \eqref{eq:guessingProb} and \eqref{eq:G} does not exceed the number of possible couples $(a,b)$ or $({\bf a},{\bf b})$. This proves proposition 1 as well as its generalization to the maximization of \eqref{eq:G} for decompositions of the form \eqref{eq:decomposition} used later in the main text.

Let us thus assume that two strategies $\lambda'$, $\lambda''$ in the considered convex decomposition of $P(ab|xy)$ are such that $\max_{a,b} \sum_{x,y,\mu} f(x,y,\mu) P_{\lambda'}(a(x,y,\mu),b(x,y,\mu)|xy)$ is achieved for the functions $a'(x,y,\mu), b'(x,y,\mu)$ and $\max_{a,b} \sum_{x,y,\mu} f(x,y,\mu) P_{\lambda''}(a(x,y,\mu), b(x,y,\mu)|xy)$ is achieved for the functions $a''(x,y,\mu)=a'(x,y,\mu)$, $b''(x,y,\mu)=b'(x,y,\mu)$. We now define a new decomposition for $P(ab|xy)=\sum_\lambda \tilde q_\lambda \tilde P_\lambda(ab|xy)$ by choosing
\begin{equation}
\tilde q_\lambda=
\begin{cases}
q_\lambda & \text{ if $\lambda \neq \lambda',\lambda''$}\\
q_{\lambda'} + q_{\lambda''} & \text{ if $\lambda = \lambda'$}\\
0 & \text{if $\lambda = \lambda''$}
\end{cases}
\end{equation}
and
\begin{equation}
\tilde P_\lambda(ab|xy)=
\begin{cases}
P_\lambda(ab|xy) & \text{ if $\lambda \neq \lambda',\lambda''$}\\
\frac{q_{\lambda'}P_{\lambda'}(ab|xy) + q_{\lambda''}P_{\lambda''}(ab|xy)}{q_{\lambda'}+q_{\lambda''}} & \text{ if $\lambda = \lambda'$}\\
0 & \text{if $\lambda = \lambda''$}.
\end{cases}
\end{equation}
Clearly, the new decomposition is a valid convex decomposition since it satisfies $\tilde q_{\lambda} \geq 0$, $\sum_\lambda\tilde q_\lambda = 1$, and $P(ab|xy)=\sum_\lambda \tilde q_\lambda \tilde P_\lambda(ab|xy)$. We are thus just left to show that it provides the same value for Eq.~\eqref{eq:proposition}. This is verified through:
\begin{equation}
\begin{split}
\sum_\lambda \tilde q_\lambda \ \underset{a,b}{\max}\ \sum_{x,y,\mu}\ f(x,y,\mu) \tilde P_\lambda(a(x,y,\mu), b(x,y,\mu)|xy) &= \sum_{\lambda\neq \lambda',\lambda''} q_\lambda \ \underset{a,b}{\max}\ \sum_{x,y,\mu} f(x,y,\mu) P_\lambda(a(x,y,\mu), b(x,y,\mu)|xy) \\
&\ \ \ \ \ \ \ \ \ \ + \tilde q_{\lambda'} \underset{a,b}{\max} \sum_{x,y,\mu}\ f(x,y,\mu) \tilde P_{\lambda'}(a(x,y,\mu), b(x,y,\mu)|xy)\\
&= \sum_\lambda q_\lambda \ \underset{a,b}{\max}\ \sum_{x,y,\mu}\ f(x,y,\mu) P_\lambda(a(x,y,\mu), b(x,y,\mu)|xy).
\end{split}
\end{equation}
%
%\medskip
%In the same way, when considering~\eqref{eq:G},
%\begin{equation*}
%G(\{q_\lambda,P_\lambda\}) = \sum_{x,y} p(x,y) G_{x,y}(\{q_\lambda,P_\lambda\})
%\end{equation*}
%it suffices to consider only a decomposition onto only as many terms as there are arguments to the optimization $(\mathbf{a}_x,\mathbf{b}_y)$, which is $(|a| |b|)^{|x| |y|}$.  The argument is the same as above; if there are more than $(|a| |b|)^{|x| |y|}$ strategies $\lambda$ in the decomposition, then a pair of extremal strategies $\lambda'$ and $\lambda''$ must exist that have $\underset{\mathbf{a}_x,\mathbf{b}_y}{\text{argmax}}\ P_{\lambda'}(\mathbf{a}_x\mathbf{b}_y|xy)=\underset{\mathbf{a}_x,\mathbf{b}_y}{\text{argmax}}\ P_{\lambda''}(\mathbf{a}_x\mathbf{b}_y|xy)$ and then, as shown above, a new decomposition can be introduced that eliminates one of the two strategies.  Therefore,  $(|a| |b|)^{|x| |y|}$ terms suffice.  In general, $(|a| |b|)^{|x| |y|}$ terms are also necessary, since if two strategies have different maximizing arguments, they cannot be combined and jointly maximized.

Thus, it suffices to consider only as many strategies $\lambda$ as the number of arguments to the maximization.
$\hfill\Box$

%%%%%%%%%%%%%%%%%%

\section{A Bell expression to certify optimal randomness extraction}
\label{sec:dual}

Here we show that whenever the primal and dual objective functions for the dual of the SDP relaxations of Eq.~\eqref{eq:bigProgram} (of which~\eqref{eq:program} is a special case) coincide, variables of the dual provide the description of a Bell expression which can be used to certify that no one can guess the outcomes observed by the parties with a probability higher than that given by the relaxation's primal.

For this, let us write the SDP relaxation of Eq.~\eqref{eq:bigProgram}, as well as its dual program~\cite{Boyd04}. For any hierarchy level, this can be done by considering for each probability distribution $P_{\balpha\bbeta}$ a matrix of the form $\Gamma_{\balpha\bbeta}=\sum_i F_i\x^i_{\balpha\bbeta}$, where $F_i$ are constant matrices and $\x^i_{\balpha\bbeta}$ are a finite number of variables~\cite{NPA07,NPA08}. For convenience, we also introduce the constant matrix $F(ab|xy)$ to pick up the terms in the $\Gamma$ matrices associated to the probabilities, i.e. $P_{\balpha\bbeta}(ab|xy)=\Tr(F(ab|xy)\Gamma_{\balpha\bbeta})$, as well as the indicative function $f_i(ab|xy) = \Tr(F(ab|xy)F_i)$. The relaxation and its dual then read:

%\jdb{--- Things updated below here ---}

\noindent
\begin{minipage}{.5\linewidth}
\begin{equation} \label{eq:relaxmax}
\begin{split}
\textit{Primal:  }\\
G(P) \leq \underset{\x^i_{\balpha\bbeta}}{\ \ \max\ \ } & \sum_{xy}p(x,y)\sum_{\balpha\bbeta i} f_i(\balpha\bbeta|xy) \x^i_{\balpha\bbeta}\\
\text{s.t.\ \ } & \sum_{\balpha\bbeta i} f_i(ab|xy) \x^i_{\balpha\bbeta} = P(ab|xy)\\
%& \sum_{ab\alpha\beta i} f_i(ab|xy)\x^i_{\alpha\beta} = 1\\
& \sum_i F_i \x^i_{\balpha\bbeta} \geq 0,\\
\ \\
\
\end{split}
\end{equation}
\end{minipage}
\begin{minipage}{.5\linewidth}
\begin{equation} \label{eq:relaxmaxdual}
\begin{split}
\textit{Dual:  }\\
\underset{c_{abxy}, M_{\balpha\bbeta}}{\ \ \min\ \ } & \sum_{abxy} c_{abxy}P(ab|xy)\\
%\text{s.t.\ \ } & \sum_{abxy} f_i(ab|xy)c_{abxy} \\
%&\ \ \ \ \ \ + \Tr(F_i M_{\balpha\bbeta}) = \sum_{xy} p(x,y) f_i(\balpha\bbeta|xy)\\
\text{s.t.\ \ } & \sum_{abxy} f_i(ab|xy)c_{abxy} + \Tr(F_i M_{\balpha\bbeta}) \\
&\ \ \ \ \ \ \ \ \ \ \ \ = \sum_{xy} p(x,y) f_i(\balpha\bbeta|xy)\\& M_{\balpha\bbeta} \leq 0.\\
\\
\end{split}
\end{equation}
\end{minipage}

Eq.~\eqref{eq:relaxmaxdual} is the dual of Eq.~\eqref{eq:relaxmax} in the sense that the value of its objective function for any set of variables $(c_{abxy},M_{\balpha\bbeta})$ satisfying its constraints sets an upper bound on the optimal value of the primal's optimization. In practice, solving these programs typically yields identical values for both objective functions, thus certifying their optimality. This was the case for all programs solved for this paper.
%In fact, non-coinciding objective functions were treated as numerical errors in this paper. We can thus restrict our attention to the case where they are equal.
%Moreover, it is known that whenever the Slater\ls{-any others?} conditions are staisfied, the optimal values of both optimizations coincide. This is indeed the case here, since one can check that the choice ... satisfies ..., and ... satisfies ... \ls{?}. The maximum value of Eq.~\eqref{eq:relaxmax} thus coincides with the minimal value of Eq.~\eqref{eq:relaxmaxdual}.

Let us now show that when the two objective functions coincide the coefficients $c_{abxy}$ define a Bell expression, whose value achieved by the tested probabilities $P(ab|xy)$, can only be achieved when the guessing probability is bounded by the optimum of the above primal. For this, we also write an SDP relaxation of equation~\eqref{eq:program2}. Bounding the maximum solution of this program for any quantum correlations achieving the value $v$ for the Bell expression defined by $c_{abxy}$ will conclude the proof.

This relaxation is done as above by associating to each probability distribution $P_{\balpha\bbeta}$ a matrix $\Gamma_{\balpha\bbeta}'=\sum_i F_i \y^i_{\balpha\bbeta}$. The obtained semi-definite program then reads:
\begin{equation}
\begin{split}
G(v)\leq \underset{\y^i_{\balpha\bbeta}}{\ \ \max\ \ } & \sum_{xy}p(x,y) \sum_{\balpha\bbeta i}f_i(\balpha\bbeta|xy) \y^i_{\balpha\bbeta}\\
\text{s.t.\ \ } & \sum_{abxy}\sum_{\balpha\bbeta i} c_{abxy} f_i(ab|xy) \y^i_{\balpha\bbeta} = v\\
& \sum_{ab\balpha\bbeta i} f_i(ab|xy) \y^i_{\balpha\bbeta} = 1\\
& \sum_i F_i \y^i_{\balpha\bbeta} \geq 0.
\end{split}
\end{equation}

So let us bound $G(v)$ for the coefficients $c_{abxy}$ of the above dual, by using the other variables of this dual at optimum. For this we write
\begin{equation}
\begin{split}
G(v) &\leq \underset{\y^i_{\balpha\bbeta i}}{\ \ \max\ \ } \sum_{xy} p(x,y) \sum_{\balpha\bbeta}f_i(\balpha\bbeta|xy) \y^i_{\balpha\bbeta}\\
& = \underset{\y^i_{\balpha\bbeta i}}{\ \ \max\ \ } \sum_{\balpha\bbeta} \left[ \sum_{abxy} f_i(ab|xy) c_{abxy} + \Tr(F_i M_{\balpha\bbeta}) \right] \y^i_{\balpha\bbeta}\\
& = \underset{\y^i_{\balpha\bbeta}}{\ \ \max\ \ } v + \sum_{\balpha\bbeta} \Tr\left( \left(\sum_i F_i \y^i_{\balpha\bbeta}\right) M_{\balpha\bbeta} \right)\\
& \leq v = \underset{\x^i_{\balpha\bbeta}}{\ \ \max\ \ } \sum_{xy} p(x,y) \sum_{\balpha\bbeta i} f_i(ab|xy) \x^i_{\balpha\bbeta},
\end{split}
\end{equation}
where we used the fact that the $M_{\balpha\bbeta}$ matrices are negative and the $\y^i_{\balpha\bbeta}$ variables must satisfy $\sum_i F_i \y^i_{\balpha\bbeta}\geq 0$ to get to the last line. This concludes the proof.
$\hfill\Box$

\section{Concavity of the guessing probability function}
\label{sec:concavity}

As mentioned in the main text, it was proposed in~\cite{PAM10} to estimate the randomness associated to a Bell inequality violation by upperbounding the guessing probability $G_{xy}(v)$ by the concave hull of the function 
\begin{equation} \label{eq:program22}
\begin{split}
f_{xy}(v) = \underset{a,b,P}{\ \ \max\ \ } & P(ab|xy)\\
\text{s.t.\ \ } & \sum_{abxy}c_{abxy}P(ab|xy) = v\\
& P(ab|xy)\text{ is quantum}.
\end{split}
\end{equation}
When relaxing in this optimization the quantum condition to some level of the NPA hierarchy, computing the function $f_{xy}(v)$ amount to solving $|a|\cdot|b|$ semidefinite programs involving a matrix of size $N$ (one optimization for each value of $(a,b)$), and taking the concave hull of the maximum of the $|a|\cdot|b|$ functions of $v$ found in this way. Let us show that taking this concave hull is not necessarily a trivial step.

For this matter, we consider the function $f_{0,0}(v)$ for the $\gamma$ inequality~\eqref{eq:newineq} with $\gamma=3/4$. The bound on this function found at local level 1 of the NPA hierarchy is plotted in figure~\ref{fig:nonConcave}. Clearly it is not concave. In particular, the value found for $v=2.4$ is $f_{00}(2.4)\simeq0.7204$, whereas its concave hull for the same $v$ takes the value $\simeq0.8075$. In fact, the result of the alternative optimization \eqref{eq:program2} presented in the main text also yields this value. Since one can check that there exists a decomposition of the form~\eqref{eq:decomposition} in terms of quantum correlations that achieve this guessing probability, it is indeed an optimal value.

\begin{figure}[h]
\includegraphics[width=0.5\textwidth]{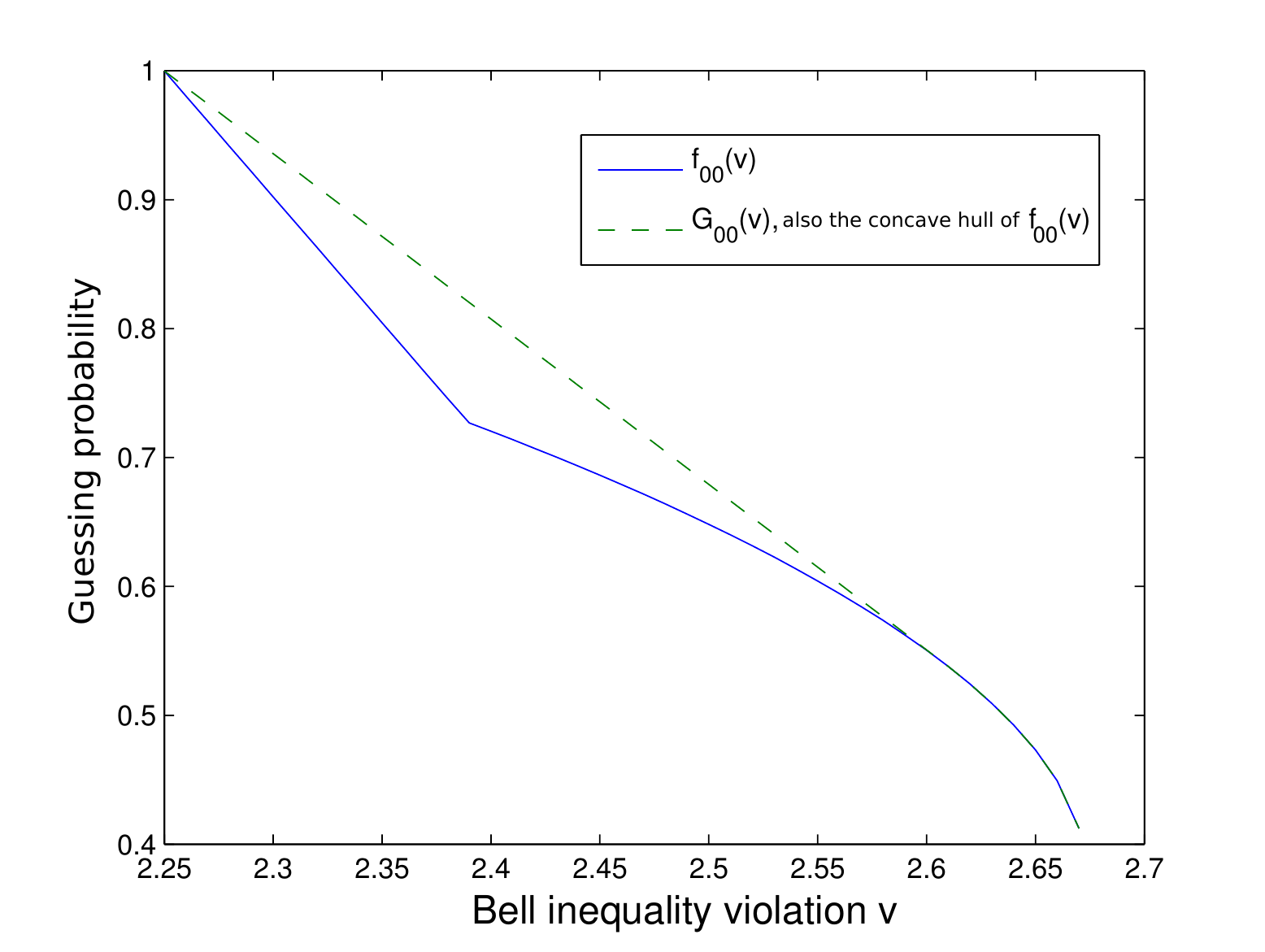}
\caption{Bounds on the guessing probability $G_{00}(v)$ found using the programs~\eqref{eq:program22} and \eqref{eq:program2} of the main text. Since the function $f_{00}(v)$ is not concave, its concave hull for $v=2.4$ requires knowledge of $f_{00}(v)$ for $v=0$ and $v\simeq 2.607$. One can check that the the function $G_{00}(v)$ coincides with the concave hull of $f_{00}(v)$.}
\label{fig:nonConcave}
\end{figure}

Finding the correct guessing probability in this case with this method thus requires one to solve \eqref{eq:program22} for both $v=0$ and $v \simeq 2.607$ (analytical value unknown), which requires some global knowledge about the $f_{00}(v)$ function. In contrast, the method presented in the main text in Eq.~\eqref{eq:program2} directly yields the guessing probability $G_{00}(2.4)$ by only considering the value $v=2.4$. It requires however, for the same level of the hierarchy, one optimization with $|a|\cdot|b|$ semidefinite blocks of size $N$ (which constitutes a larger problem than the individual ones above).

%%%%%%%%%%%%%%%%%%

\section{Application to~[Christensen \textit{et al.}, Phys. Rev. Lett. 111, 130406 (2013)]}
\label{sec:exptAnalysis}

Nonlocal correlations observed with limited detection efficiency (about $75\%$) are reported in~\cite{Christensen13}\footnote{Note that a similar claim is presented in~\cite{Giustina13}, but without enough details to allow for a quantification of the randomness created during the experiment. We thus restrict our analysis to the data reported in~\cite{Christensen13}.}. An estimation of randomness based on these observations is then also provided, with the help of known theoretical techniques. Here we perform a similar analysis by using the method described in the main text. Namely we estimate the amount of private randomness one could expect to extract from the setup used for this experiment in the limit of many runs.

%\jdb{First of all, we need to slightly process the data reported in~\cite{Christensen13} in order to make it fit our theoretical description. Indeed, since only a finite number of measurements were performed in~\cite{Christensen13},} the frequencies with which the different outcomes were observed violate the no-signalling conditions, i.e.

Before applying our tools to this data, we need to slightly process it in order to make it fit with our theoretical description. Indeed, being the result of a finite number of measurements, the frequencies with which the different outcomes were observed violate the no-signalling conditions, i.e.
%Indeed, we assumed in the main text to be in the limit in which infinitely many measurements were performed, but only a finite number of measurements were performed in~\cite{Christensen13}. As a consequence, the frequencies with which the different outcomes were observed violate the no-signalling conditions, i.e.
\begin{equation}
\frac{S_A(a,b)}{N(a,b)} \neq \frac{S_A(a,b')}{N(a,b')}
\end{equation}
where $S_A$ is the number of singles observed at Alice, and $N$ the number of events recorded in presence of some settings (we use here the notation of~\cite{Christensen13}). All quantum correlations being no-signalling, this violation prevents us from applying the optimizations presented in the main text to these frequencies directly.

Since one expects this signalling character to disappear in the limit of many measurements, we choose to remove this feature by projecting the frequencies onto the no-signalling probability subspace in an orthogonal manner.
%\jdb{applying to the observed frequencies an orthogonal projection onto the no-signalling probability subspace}.
This can be done by considering the full probability space as $\mathbb{R}^{16}$ equipped with the standard inner product. Alternatively, this projection can be performed directly in terms of expectation value for the $\pm 1$ eigenvalue observables $A_{xy}$, $B_{xy}$, as we do now: first we compute the expectation values for each setting $(x,y)$ directly from the counts as
\begin{equation}
\langle A_{00} \rangle = \frac{2S_A(a,b)-N(a,b)}{N(a,b)},\ \ 
\langle B_{00} \rangle = \frac{2S_B(a,b)-N(a,b)}{N(a,b)},\ \ 
\langle A_{00} B_{00} \rangle = \frac{4C(a,b) -2S_A(a,b) -2S_B(a,b) + N(a,b)}{N(a,b)}
\end{equation}
and similarly for the other pairs of settings. The projection is then obtained by setting $\langle A_x \rangle = \frac{\langle A_{x0} \rangle + \langle A_{x1} \rangle}{2}$ and $\langle B_y \rangle = \frac{\langle B_{0y} \rangle + \langle B_{1y} \rangle}{2}$.

The no-signalling correlations obtained in this way take value:
\begin{equation}\label{eq:nosigCorrs}
\begin{split}
\langle A_0 \rangle = -0.9966077,&\ \ \ \ \ \langle A_1 \rangle = -0.9891902\\
\langle B_0 \rangle = -0.9965994,&\ \ \ \ \ \langle B_1 \rangle = -0.9896845\\
\langle A_{00} B_{00} \rangle = 0.9975133,&\ \ \ \ \ \langle A_{01} B_{01} \rangle = 0.9911125\\
\langle A_{10} B_{10} \rangle = 0.9907040,&\ \ \ \ \ \langle A_{11} B_{11} \rangle = 0.9791706,
\end{split}
\end{equation}
yielding a CHSH violation of $\sim 2.000159$.

Note that the correlations \eqref{eq:nosigCorrs} could still, in principle, lie outside the set of quantum correlations, thus still preventing a direct application of our technique. We checked however that this was not the case.

In general, other projections could also be chosen to relate the observed signalling statistics to some quantum correlations. By construction, the particular projection used here minimizes the euclidean distance between the projected probabilities and the original ones. This ensures that its result cannot be too far away from the original statistics, as measured by any $p$-norm. In particular, it must be close to any point belonging both to the quantum set and to a confidence ball of radius $\epsilon$ in $p$-norm around the original probabilities.

Running the programs~\eqref{eq:bigProgram} and~\eqref{eq:program2} either with the obtained correlations~\eqref{eq:nosigCorrs} or the CHSH violation, and a uniform choice of settings ($p(x,y)=1/4$) certifies in both cases a private randomness rate of 0.00014567 [bits per run] $\simeq$ 16207 [bits] / 111259682 [runs]. %Considering that 111259682 measurements were performed, this amounts to 16207 bits. 
This is about twice as much randomness per run as reported in~\cite{Christensen13}\footnote{Note that in~\cite{Christensen13} authors consider extracting randomness from the outcomes observed by one party rather than from joint outcomes. For the small CHSH violation observed here, the amount of randomness present in the joint outcomes is however sensibly similar (see~\cite{PAM10}).}.
%(compare to 8700 bits obtained from the marginal in~\cite{Christensen13}.).

%%%%%%%%%%%%%%%%%%

\section{Properties of an Inequality for Randomness Certification}
\label{sec:gammaineq}

Let us provide a brief description of the properties of inequality~\eqref{eq:newineq}. For $\gamma > 1$, to increase the value that this expression takes the first correlation term should be made larger (in absolute value) at the expense of the other terms.  To do this, measurements $A_0$ and $B_0$ should be chosen to have an inner product smaller than $\frac{1}{\sqrt{2}}$.  This expression reaches its maximum value for the maximally entangled states $\ket{\Phi^{+}}$, $\ket{\Psi^{-}}$ and the following set of coplanar measurement angles:
\begin{eqnarray}
\mathbf{a_0} . \mathbf{b_0} &=& \cos \theta_{00}  \nonumber \\
\mathbf{a_0} . \mathbf{b_1} &=& \cos \alpha    \label{eq:measgamma}    \\
\mathbf{a_1} . \mathbf{b_0} &=& \cos \alpha     \nonumber \\
\mathbf{a_1} . \mathbf{b_1} &=& \cos (\pi-\alpha)  \nonumber
\end{eqnarray}
where $\mathbf{a_i}, \mathbf{b_i}$ are coplanar vectors specifying the measurement settings $A_i,B_i$, and $\alpha = \frac{\pi-\theta_{00}}{3}$.  The angle $\theta_{00}$ can be expressed in terms of $\gamma$ as:
\begin{equation}
\theta_{00}(\gamma) = 3 \cos^{-1}\left(\frac{\sqrt{5 + (1/\gamma)\big(\sqrt{3}\sqrt{(3\gamma -1)(\gamma+1)} - 1} \big)}{2\sqrt{2}}\right) \ .
\end{equation}
The maximum quantum value $V_{Q}$ of the inequality also depends on $\gamma$:
\begin{equation}
V_{Q}(\gamma) = \gamma \cos\left(\theta_{00}(\gamma)\right) + 3 \sin \left(\frac{\pi}{6} + \frac{\theta_{00}(\gamma)}{3} \right)
\end{equation}	
which is found from computing $\Tr(V \ket{\Psi^-}\bra{\Psi^-})$.

%\ls{Add plots of randomness as a function of $\eta$, $\gamma$ as a function of $\eta$ (check that optimization over all settings are not better than CHSH settings, and check what happens with a third measurement (!)).}

Let us now consider how to find a bound on the amount of marginal randomness that a particular value $v$ of the Bell inequality~\eqref{eq:newineq} implies. Here, by marginal randomness we mean the randomness of only one party's outcome assuming that the other party's outcome is securely destroyed. Following~\cite{PAM10,PhysRevLett.108.100402}, we are interested in calculating the guessing probability of the most biased pure state for which $\Tr(V \rho) = v$.  The family of biased states we require here are of the form $\ket{\Phi(\beta)} = \cos \beta \ket{00} + \sin \beta \ket{11}$.

The maximum value that this Bell inequality can take using a state of the form $\ket{\Phi(\beta)}$ is found by optimizing the measurements, $A_0, A_1, B_0, B_1$ in:
\begin{equation}
\hat{V} =   A_0 (\gamma B_0 + B_1) +  A_1 (B_0 - B_1) \,.
\end{equation}
The optimal violation is found when $\mathbf{a_0} = \frac{\gamma \mathbf{b_0} - \mathbf{b_1}}{|\gamma \mathbf{b_0} - \mathbf{b_1}|}$ and $\mathbf{a_1} = \frac{\mathbf{b_0} - \mathbf{b_1}}{|\mathbf{b_0} - \mathbf{b_1}|}$.  

Using this observation about the optimal alignment of the measurement operators $A_0$ and $A_1$ and following the approach of~\cite{HHH95}, we can conclude that 
\begin{equation}
V_{Q}(\Phi(\beta)) =   \max_{v_0, v_1} \| M \mathbf{v}_0 \| +  \| M \mathbf{v}_1 \|
\end{equation}
where 
\begin{equation}
\mathbf{v}_0 = \left( \begin{array}{c}
\gamma b_x^0 + b_x^1 \\
0 \\
\gamma b_z^0 + b_z^1
\end{array} \right) \, , \text{ and } 
\mathbf{v}_1 = \left( \begin{array}{c}
b_x^0 - b_x^1 \\
0 \\
b_z^0 - b_z^1
\end{array} \right) \, ,
\end{equation}
(a symmetry allows the $y$ values to be chosen as $0$) and $M$ is the matrix representing the state $\ket{\Phi(\beta)}$ on the Bloch sphere.  $M$ is diagonal with eigenvalues $\sin(2\beta),-\sin(2\beta),1$.  Then setting $b_x^0 = \cos\phi_0$ and $b_x^1 = \cos\phi_1$,
\begin{align}
V_{Q}(\Phi(\beta)) =   \max_{\phi_0, \phi_1} & \left[ \sqrt{(\gamma \cos \phi_0 + \cos \phi_1)^2 + \sin^2(2\beta)(\gamma \sin \phi_0 + \sin \phi_1)^2} \right.  \nonumber \\
& \ \ \left. + \, \sqrt{(\cos \phi_0 - \cos \phi_1)^2 + \sin^2(2\beta)(\sin \phi_0 - \sin \phi_1)^2} \right] \, .
\end{align}
We cannot solve this expression analytically for general $\gamma$, however for cases of practical interest, $\gamma$ will be only a little larger than one.  

In order to obtain an analytic estimate of the randomness, we assume that the values of $\gamma$ lie in $1\leq\gamma\lesssim 1.1$. Let us assume that within this interval the expression for the maximum violation can be expanded around $\gamma = 1$ as
\begin{eqnarray}
V_{Q}(\Phi(\beta)) &=&  2\sqrt{1+\sin^2(2\beta)} + \frac{\epsilon}{\sqrt{1+\sin^2(2\beta)}} + \frac{\epsilon^2}{(1+\sin^2(2\beta))^{3/2}} \left(\frac{\sin^2(2\beta)}{2}+x\right) + O(\epsilon^3) \, ,
\end{eqnarray}
where we set $\epsilon = \gamma-1$, $x:=\sin^2(2\beta)(\sin^2(2\beta) \phi_0'[1] - \phi_1'[1])-(\phi_0'[1]-\sin^2(2\beta)\phi_1'[1])^2$ and $\phi_0'[1] := \left. \frac{\partial \phi_0}{\partial\gamma}\right|_{\gamma=1}$ and likewise for $\phi_1'[1]$.  In order to upper bound $V_{Q}(\Phi(\beta))$, we require a bound on $x$.  Expanding $\frac{dV_Q}{d\gamma}$ in terms of partial differentials and equating the first-order in $\epsilon$ term on both sides gives $\phi_0'[1] = \frac{1}{2} \sin^2(2\beta) (1 + 2\phi_1'[1])$.  Further, since we have expanded $\cos(\phi_1)$ in the neighbourhood of $\gamma = 1$ as $\cos(\pi/2 + \epsilon \phi_1'[1])$, we should have $-\pi < \epsilon \phi_1'[1] < \pi$, since we can restrict $\phi_1$ to occupy only a range over $2\pi$.  Note that this is equivalent to making the assumption that the expansion is valid for this range of $\gamma$.  Then 
\begin{eqnarray}
\epsilon x &\leq& \frac{\epsilon}{(1+\sin^2(2\beta))^{3/2}} \left(\pi \sin^4(2\beta) (1-\sin^2(2\beta)) \phantom{\frac{1}{2}}  +  \frac{\epsilon}{2} \sin^2(2\beta) \left(1+\sin^4(2\beta) - \frac{1}{2}\sin^2(2\beta)  \right) \right) \, .  
\end{eqnarray}
(Note that using the analytic bound, when $\beta =\pi/4$, $x=1/4$, and from a numerical analysis, $x < 1.6$ for all values of $\beta$.)  Then, we can supply a bound on the maximum value that the inequality~\eqref{eq:newineq} can take for any value of $\beta$ and small values of $(\gamma -1)$ as:
\begin{eqnarray}
V_{Q}(\Phi(\beta)) &\leq&  2\sqrt{1+\sin^2(2\beta)}  + \frac{\epsilon}{\sqrt{1+\sin^2(2\beta)}} \left(1+ \frac{\pi \sin^4(2\beta)}{1+\sin^2(2\beta)}\bigl( 1-\sin^2(2\beta)  \bigr)  \right)   \nonumber \\
 & & \qquad + \,  \frac{\epsilon^2 \sin^2(2\beta)}{2(1+\sin^2(2\beta))^{3/2}}  \left(1+\sin^4(2\beta) - \frac{1}{2}\sin^2(2\beta)  \right)  +  O(\epsilon^3) \, ,
\label{eq:Vbound}
\end{eqnarray}

Under the assumption that the expansion is valid in the considered range (this is equivalent to assuming that the derivatives of $V$ with respect to $\gamma$ are not too large), this method gives a bound on the randomness that a particular observed violation implies, and thus a quick and simple method to calculate a lower bound on the extractable intrinsic randomness of an experiment.  Unfortunately, one can check that this quick bound does not reproduce the value obtained numerically using programs presented in the main text, and in particular it does not improve on the bound implied by the CHSH inequality, at least not for values of $\gamma$ in the range of interest.

The reason for this is as follows.  One way to bound the intrinsic min-entropy of one party's outcome, after observing a value of the inequality~(\ref{eq:newineq}) $v$, is to find the largest value of $\beta$ compatible with $v$ such that $V_Q = v$ using the bound~(\ref{eq:Vbound}).  Then the guessing probability for any choice of input measurement is $G \leq \cos^2 \beta$ for $0\leq\beta\leq\frac{\pi}{4}$.  However, it is not sufficient to use that bound, because it returns a lower amount of marginal intrinsic randomness than does the same technique using instead the CHSH inequality.  In actual fact, the amount of generated randomness is always \emph{higher} than $\cos^2 \beta$ because the $\gamma$ factor in the inequality breaks symmetry in the system, meaning that the orientation of the state with respect to the measurements does not allow either of Alice or Bob's measurements to be along the $z$-direction, and achieve the value $v$ with the state $\ket{\Phi(\beta)}$.  If none of the measurements are along the $z$-axis then the marginal randomness must be greater than $\cos^2 \beta$.  Therefore, it is necessary to optimize the randomness over the state angle $\beta$ and the angle of orientation of Alice's measurement to the $z$-axis, in order to improve on the bound given in~\cite{PAM10} for the CHSH inequality. However this is not taken into account in the presented analysis.

%%%%%%%%%%%%%%%%%%

\bibliographystyle{unsrt}	
\bibliography{randomness3}

\begin{thebibliography}{30}
\expandafter\ifx\csname natexlab\endcsname\relax\def\natexlab#1{#1}\fi
\expandafter\ifx\csname bibnamefont\endcsname\relax
  \def\bibnamefont#1{#1}\fi
\expandafter\ifx\csname bibfnamefont\endcsname\relax
  \def\bibfnamefont#1{#1}\fi
\expandafter\ifx\csname citenamefont\endcsname\relax
  \def\citenamefont#1{#1}\fi
\expandafter\ifx\csname url\endcsname\relax
  \def\url#1{\texttt{#1}}\fi
\expandafter\ifx\csname urlprefix\endcsname\relax\def\urlprefix{URL }\fi
\providecommand{\bibinfo}[2]{#2}
\providecommand{\eprint}[2][]{\url{#2}}

\bibitem[{\citenamefont{Vadhan}(2012)}]{Vadhan}
\bibinfo{author}{\bibfnamefont{S.}~\bibnamefont{Vadhan}},
  \emph{\bibinfo{title}{Pseudorandomness}}, vol.~\bibinfo{volume}{7} of
  \emph{\bibinfo{series}{Foundations and Trends in Theoretical Computer
  Science}} (\bibinfo{publisher}{Now Publishers}, \bibinfo{year}{2012}).

\bibitem[{\citenamefont{Colbeck}(2006)}]{Colbeckthesis}
\bibinfo{author}{\bibfnamefont{R.}~\bibnamefont{Colbeck}}, Ph.D. thesis,
  \bibinfo{school}{University of Cambridge} (\bibinfo{year}{2006}),
  \bibinfo{note}{arXiv:0911.3814}.

\bibitem[{\citenamefont{Pironio et~al.}(2010)\citenamefont{Pironio, Ac\'{i}n,
  Massar, de~la Giroday, Matsukevich, Maunz, Olmschenk, Hayes, Luo, Manning
  et~al.}}]{PAM10}
\bibinfo{author}{\bibfnamefont{S.}~\bibnamefont{Pironio}},
  \bibinfo{author}{\bibfnamefont{A.}~\bibnamefont{Ac\'{i}n}},
  \bibinfo{author}{\bibfnamefont{S.}~\bibnamefont{Massar}},
  \bibinfo{author}{\bibfnamefont{A.~B.} \bibnamefont{de~la Giroday}},
  \bibinfo{author}{\bibfnamefont{D.~N.} \bibnamefont{Matsukevich}},
  \bibinfo{author}{\bibfnamefont{P.}~\bibnamefont{Maunz}},
  \bibinfo{author}{\bibfnamefont{S.}~\bibnamefont{Olmschenk}},
  \bibinfo{author}{\bibfnamefont{D.}~\bibnamefont{Hayes}},
  \bibinfo{author}{\bibfnamefont{L.}~\bibnamefont{Luo}},
  \bibinfo{author}{\bibfnamefont{T.~A.} \bibnamefont{Manning}},
  \bibnamefont{et~al.}, \bibinfo{journal}{Nature}
  \textbf{\bibinfo{volume}{464}}, \bibinfo{pages}{1021} (\bibinfo{year}{2010}).

\bibitem[{\citenamefont{Pironio and Massar}(2013)}]{PM11}
\bibinfo{author}{\bibfnamefont{S.}~\bibnamefont{Pironio}} \bibnamefont{and}
  \bibinfo{author}{\bibfnamefont{S.}~\bibnamefont{Massar}},
  \bibinfo{journal}{Phys. Rev. A} \textbf{\bibinfo{volume}{87}},
  \bibinfo{pages}{012336} (\bibinfo{year}{2013}).

\bibitem[{\citenamefont{Fehr et~al.}(2013)\citenamefont{Fehr, Gelles, and
  Schaffner}}]{FGS11}
\bibinfo{author}{\bibfnamefont{S.}~\bibnamefont{Fehr}},
  \bibinfo{author}{\bibfnamefont{R.}~\bibnamefont{Gelles}}, \bibnamefont{and}
  \bibinfo{author}{\bibfnamefont{C.}~\bibnamefont{Schaffner}},
  \bibinfo{journal}{Phys. Rev. A} \textbf{\bibinfo{volume}{87}},
  \bibinfo{pages}{012335} (\bibinfo{year}{2013}).

\bibitem[{\citenamefont{Vazirani and Vidick}(2012)}]{VV12}
\bibinfo{author}{\bibfnamefont{U.}~\bibnamefont{Vazirani}} \bibnamefont{and}
  \bibinfo{author}{\bibfnamefont{T.}~\bibnamefont{Vidick}}
  (\bibinfo{year}{2012}), \bibinfo{note}{arXiv:1111.6054}.

\bibitem[{\citenamefont{Colbeck and Renner}(2012)}]{CR12}
\bibinfo{author}{\bibfnamefont{R.}~\bibnamefont{Colbeck}} \bibnamefont{and}
  \bibinfo{author}{\bibfnamefont{R.}~\bibnamefont{Renner}},
  \bibinfo{journal}{Nature Physics} \textbf{\bibinfo{volume}{8}},
  \bibinfo{pages}{450} (\bibinfo{year}{2012}).

\bibitem[{\citenamefont{Gallego et~al.}(2013)\citenamefont{Gallego, Masanes,
  Torre, Dhara, Aolita, and Ac\'{i}n}}]{GMTDAA12}
\bibinfo{author}{\bibfnamefont{R.}~\bibnamefont{Gallego}},
  \bibinfo{author}{\bibfnamefont{L.}~\bibnamefont{Masanes}},
  \bibinfo{author}{\bibfnamefont{G.~D.~L.} \bibnamefont{Torre}},
  \bibinfo{author}{\bibfnamefont{C.}~\bibnamefont{Dhara}},
  \bibinfo{author}{\bibfnamefont{L.}~\bibnamefont{Aolita}}, \bibnamefont{and}
  \bibinfo{author}{\bibfnamefont{A.}~\bibnamefont{Ac\'{i}n}},
  \bibinfo{journal}{Nat. Commun.} \textbf{\bibinfo{volume}{4}},
  \bibinfo{pages}{2654} (\bibinfo{year}{2013}).

\bibitem[{\citenamefont{Mironowicz and Paw{\l}owski}(2013)}]{MP13}
\bibinfo{author}{\bibfnamefont{P.}~\bibnamefont{Mironowicz}} \bibnamefont{and}
  \bibinfo{author}{\bibfnamefont{M.}~\bibnamefont{Paw{\l}owski}}
  (\bibinfo{year}{2013}), \bibinfo{note}{arXiv:1301.7722}.

\bibitem[{\citenamefont{Mironowicz and Pawlowski}(2013)}]{Pawlowski13}
\bibinfo{author}{\bibfnamefont{P.}~\bibnamefont{Mironowicz}} \bibnamefont{and}
  \bibinfo{author}{\bibfnamefont{M.}~\bibnamefont{Pawlowski}},
  \bibinfo{journal}{Phys. Rev. A} \textbf{\bibinfo{volume}{88}},
  \bibinfo{pages}{032319} (\bibinfo{year}{2013}).

\bibitem[{\citenamefont{Christensen et~al.}(2013)\citenamefont{Christensen,
  McCusker, Altepeter, Calkins, Gerrits, Lita, Miller, Shalm, Zhang, Nam
  et~al.}}]{Christensen13}
\bibinfo{author}{\bibfnamefont{B.~G.} \bibnamefont{Christensen}},
  \bibinfo{author}{\bibfnamefont{K.~T.} \bibnamefont{McCusker}},
  \bibinfo{author}{\bibfnamefont{J.}~\bibnamefont{Altepeter}},
  \bibinfo{author}{\bibfnamefont{B.}~\bibnamefont{Calkins}},
  \bibinfo{author}{\bibfnamefont{T.}~\bibnamefont{Gerrits}},
  \bibinfo{author}{\bibfnamefont{A.}~\bibnamefont{Lita}},
  \bibinfo{author}{\bibfnamefont{A.}~\bibnamefont{Miller}},
  \bibinfo{author}{\bibfnamefont{L.~K.} \bibnamefont{Shalm}},
  \bibinfo{author}{\bibfnamefont{Y.}~\bibnamefont{Zhang}},
  \bibinfo{author}{\bibfnamefont{S.~W.} \bibnamefont{Nam}},
  \bibnamefont{et~al.}, \bibinfo{journal}{Phys. Rev. Lett.}
  \textbf{\bibinfo{volume}{111}}, \bibinfo{pages}{130406}
  (\bibinfo{year}{2013}).

\bibitem[{\citenamefont{Hall}(2011)}]{H11}
\bibinfo{author}{\bibfnamefont{M.~J.~W.} \bibnamefont{Hall}},
  \bibinfo{journal}{Phys. Rev. A} \textbf{\bibinfo{volume}{84}},
  \bibinfo{pages}{022102} (\bibinfo{year}{2011}).

\bibitem[{\citenamefont{Barrett and Gisin}(2011)}]{BG11}
\bibinfo{author}{\bibfnamefont{J.}~\bibnamefont{Barrett}} \bibnamefont{and}
  \bibinfo{author}{\bibfnamefont{N.}~\bibnamefont{Gisin}},
  \bibinfo{journal}{Phys. Rev. Lett.} \textbf{\bibinfo{volume}{106}},
  \bibinfo{pages}{100406} (\bibinfo{year}{2011}).

\bibitem[{\citenamefont{Koh et~al.}(2012)\citenamefont{Koh, Hall, Setiawan,
  Pope, Marletto, Kay, Scarani, and Ekert}}]{KHSPMKSE12}
\bibinfo{author}{\bibfnamefont{D.~E.} \bibnamefont{Koh}},
  \bibinfo{author}{\bibfnamefont{M.~J.~W.} \bibnamefont{Hall}},
  \bibinfo{author}{\bibnamefont{Setiawan}},
  \bibinfo{author}{\bibfnamefont{J.~E.} \bibnamefont{Pope}},
  \bibinfo{author}{\bibfnamefont{C.}~\bibnamefont{Marletto}},
  \bibinfo{author}{\bibfnamefont{A.}~\bibnamefont{Kay}},
  \bibinfo{author}{\bibfnamefont{V.}~\bibnamefont{Scarani}}, \bibnamefont{and}
  \bibinfo{author}{\bibfnamefont{A.}~\bibnamefont{Ekert}},
  \bibinfo{journal}{Phys. Rev. Lett.} \textbf{\bibinfo{volume}{109}},
  \bibinfo{pages}{160404} (\bibinfo{year}{2012}).

\bibitem[{\citenamefont{Le et~al.}(2013)\citenamefont{Le, Sheridan, and
  Scarani}}]{LSS13}
\bibinfo{author}{\bibfnamefont{T.}~\bibnamefont{Le}},
  \bibinfo{author}{\bibfnamefont{L.}~\bibnamefont{Sheridan}}, \bibnamefont{and}
  \bibinfo{author}{\bibfnamefont{V.}~\bibnamefont{Scarani}},
  \bibinfo{journal}{Phys. Rev. A} \textbf{\bibinfo{volume}{87}},
  \bibinfo{pages}{062121} (\bibinfo{year}{2013}).

\bibitem[{\citenamefont{Gill}(2003)}]{Gill03}
\bibinfo{author}{\bibfnamefont{R.}~\bibnamefont{Gill}},
  \bibinfo{journal}{Mathematical Statistics and Applications: Festschrift for
  Constance van Eeden. Eds: M. Moore, S. Froda and C. L\'eger. IMS Lecture
  Notes -- Monograph Series} \textbf{\bibinfo{volume}{42}},
  \bibinfo{pages}{133} (\bibinfo{year}{2003}).

\bibitem[{\citenamefont{Barrett et~al.}(2002)\citenamefont{Barrett, Collins,
  Hardy, Kent, and Popescu}}]{BCHK+02}
\bibinfo{author}{\bibfnamefont{J.}~\bibnamefont{Barrett}},
  \bibinfo{author}{\bibfnamefont{D.}~\bibnamefont{Collins}},
  \bibinfo{author}{\bibfnamefont{L.}~\bibnamefont{Hardy}},
  \bibinfo{author}{\bibfnamefont{A.}~\bibnamefont{Kent}}, \bibnamefont{and}
  \bibinfo{author}{\bibfnamefont{S.}~\bibnamefont{Popescu}},
  \bibinfo{journal}{Phys. Rev. A} \textbf{\bibinfo{volume}{66}},
  \bibinfo{pages}{042111} (\bibinfo{year}{2002}).

\bibitem[{\citenamefont{Ac\'in et~al.}(2012)\citenamefont{Ac\'in, Massar, and
  Pironio}}]{PhysRevLett.108.100402}
\bibinfo{author}{\bibfnamefont{A.}~\bibnamefont{Ac\'in}},
  \bibinfo{author}{\bibfnamefont{S.}~\bibnamefont{Massar}}, \bibnamefont{and}
  \bibinfo{author}{\bibfnamefont{S.}~\bibnamefont{Pironio}},
  \bibinfo{journal}{Phys. Rev. Lett.} \textbf{\bibinfo{volume}{108}},
  \bibinfo{pages}{100402} (\bibinfo{year}{2012}).

\bibitem[{\citenamefont{Dhara et~al.}(2013)\citenamefont{Dhara, de~la Torre,
  and Ac\'{i}n}}]{arXiv:1305.5384}
\bibinfo{author}{\bibfnamefont{C.}~\bibnamefont{Dhara}},
  \bibinfo{author}{\bibfnamefont{G.}~\bibnamefont{de~la Torre}},
  \bibnamefont{and} \bibinfo{author}{\bibfnamefont{A.}~\bibnamefont{Ac\'{i}n}},
  \bibinfo{journal}{Phys. Rev. Lett.} \textbf{\bibinfo{volume}{112}},
  \bibinfo{pages}{100402} (\bibinfo{year}{2013}).

\bibitem[{\citenamefont{Navascues et~al.}(2007)\citenamefont{Navascues,
  Pironio, and Acin}}]{NPA07}
\bibinfo{author}{\bibfnamefont{M.}~\bibnamefont{Navascues}},
  \bibinfo{author}{\bibfnamefont{S.}~\bibnamefont{Pironio}}, \bibnamefont{and}
  \bibinfo{author}{\bibfnamefont{A.}~\bibnamefont{Acin}},
  \bibinfo{journal}{Phys. Rev. Lett.} \textbf{\bibinfo{volume}{98}},
  \bibinfo{pages}{010401} (\bibinfo{year}{2007}).

\bibitem[{\citenamefont{Navascues et~al.}(2008)\citenamefont{Navascues,
  Pironio, and Acin}}]{NPA08}
\bibinfo{author}{\bibfnamefont{M.}~\bibnamefont{Navascues}},
  \bibinfo{author}{\bibfnamefont{S.}~\bibnamefont{Pironio}}, \bibnamefont{and}
  \bibinfo{author}{\bibfnamefont{A.}~\bibnamefont{Acin}}, \bibinfo{journal}{New
  Journal of Physics} \textbf{\bibinfo{volume}{10}}, \bibinfo{pages}{073013}
  (\bibinfo{year}{2008}).

\bibitem[{\citenamefont{Gerhardt et~al.}(2011)\citenamefont{Gerhardt, Liu,
  Lamas-Linares, Skaar, Scarani, Makarov, and
  Kurtsiefer}}]{PhysRevLett.107.170404}
\bibinfo{author}{\bibfnamefont{I.}~\bibnamefont{Gerhardt}},
  \bibinfo{author}{\bibfnamefont{Q.}~\bibnamefont{Liu}},
  \bibinfo{author}{\bibfnamefont{A.}~\bibnamefont{Lamas-Linares}},
  \bibinfo{author}{\bibfnamefont{J.}~\bibnamefont{Skaar}},
  \bibinfo{author}{\bibfnamefont{V.}~\bibnamefont{Scarani}},
  \bibinfo{author}{\bibfnamefont{V.}~\bibnamefont{Makarov}}, \bibnamefont{and}
  \bibinfo{author}{\bibfnamefont{C.}~\bibnamefont{Kurtsiefer}},
  \bibinfo{journal}{Phys. Rev. Lett.} \textbf{\bibinfo{volume}{107}},
  \bibinfo{pages}{170404} (\bibinfo{year}{2011}).

\bibitem[{\citenamefont{Giustina et~al.}(2013)\citenamefont{Giustina, Mech,
  Ramelow, Wittmann, Kofler, Beyer, Lita, Calkins, Gerrits, Nam
  et~al.}}]{Giustina13}
\bibinfo{author}{\bibfnamefont{M.}~\bibnamefont{Giustina}},
  \bibinfo{author}{\bibfnamefont{A.}~\bibnamefont{Mech}},
  \bibinfo{author}{\bibfnamefont{S.}~\bibnamefont{Ramelow}},
  \bibinfo{author}{\bibfnamefont{B.}~\bibnamefont{Wittmann}},
  \bibinfo{author}{\bibfnamefont{J.}~\bibnamefont{Kofler}},
  \bibinfo{author}{\bibfnamefont{J.}~\bibnamefont{Beyer}},
  \bibinfo{author}{\bibfnamefont{A.}~\bibnamefont{Lita}},
  \bibinfo{author}{\bibfnamefont{B.}~\bibnamefont{Calkins}},
  \bibinfo{author}{\bibfnamefont{T.}~\bibnamefont{Gerrits}},
  \bibinfo{author}{\bibfnamefont{S.~W.} \bibnamefont{Nam}},
  \bibnamefont{et~al.}, \bibinfo{journal}{Nature}
  \textbf{\bibinfo{volume}{497}}, \bibinfo{pages}{227} (\bibinfo{year}{2013}).

\bibitem[{\citenamefont{Eberhardt}(1993)}]{E93}
\bibinfo{author}{\bibfnamefont{P.~H.} \bibnamefont{Eberhardt}},
  \bibinfo{journal}{Phys. Rev. A} \textbf{\bibinfo{volume}{47}},
  \bibinfo{pages}{R747} (\bibinfo{year}{1993}).

\bibitem[{\citenamefont{Sturm}(1999)}]{S98guide}
\bibinfo{author}{\bibfnamefont{J.}~\bibnamefont{Sturm}},
  \bibinfo{journal}{Optimization Methods and Software}
  \textbf{\bibinfo{volume}{11--12}}, \bibinfo{pages}{625}
  (\bibinfo{year}{1999}), \bibinfo{note}{version 1.05},
  \urlprefix\url{http://fewcal.kub.nl/sturm}.

\bibitem[{\citenamefont{Moroder et~al.}(2013)\citenamefont{Moroder, Bancal,
  Liang, Hofmann, and G\"uhne}}]{Moroder13}
\bibinfo{author}{\bibfnamefont{T.}~\bibnamefont{Moroder}},
  \bibinfo{author}{\bibfnamefont{J.-D.} \bibnamefont{Bancal}},
  \bibinfo{author}{\bibfnamefont{Y.-C.} \bibnamefont{Liang}},
  \bibinfo{author}{\bibfnamefont{M.}~\bibnamefont{Hofmann}}, \bibnamefont{and}
  \bibinfo{author}{\bibfnamefont{O.}~\bibnamefont{G\"uhne}},
  \bibinfo{journal}{Phys. Rev. Lett.} \textbf{\bibinfo{volume}{111}},
  \bibinfo{pages}{030501} (\bibinfo{year}{2013}).

\bibitem[{faa()}]{faacets.com}
\urlprefix\url{http://www.faacets.com/db/solved}.

\bibitem[{\citenamefont{Nieto-Silleras
  et~al.}(2014)\citenamefont{Nieto-Silleras, Pironio, and Silman}}]{Silleras13}
\bibinfo{author}{\bibfnamefont{O.}~\bibnamefont{Nieto-Silleras}},
  \bibinfo{author}{\bibfnamefont{S.}~\bibnamefont{Pironio}}, \bibnamefont{and}
  \bibinfo{author}{\bibfnamefont{J.}~\bibnamefont{Silman}},
  \bibinfo{journal}{New. J. Phys.} \textbf{\bibinfo{volume}{16}},
  \bibinfo{pages}{013035} (\bibinfo{year}{2014}).

\bibitem[{\citenamefont{Boyd and Vandenberghe}(2004)}]{Boyd04}
\bibinfo{author}{\bibfnamefont{S.}~\bibnamefont{Boyd}} \bibnamefont{and}
  \bibinfo{author}{\bibfnamefont{S.}~\bibnamefont{Vandenberghe}},
  \emph{\bibinfo{title}{Convex optimization}} (\bibinfo{publisher}{Cambridge
  University Press}, \bibinfo{year}{2004}).

\bibitem[{\citenamefont{Horodecki et~al.}(1995)\citenamefont{Horodecki,
  Horodecki, and Horodecki}}]{HHH95}
\bibinfo{author}{\bibfnamefont{R.}~\bibnamefont{Horodecki}},
  \bibinfo{author}{\bibfnamefont{P.}~\bibnamefont{Horodecki}},
  \bibnamefont{and}
  \bibinfo{author}{\bibfnamefont{M.}~\bibnamefont{Horodecki}},
  \bibinfo{journal}{Phys. Lett. A} \textbf{\bibinfo{volume}{200}},
  \bibinfo{pages}{340} (\bibinfo{year}{1995}).

\end{thebibliography}

\end{widetext}

\end{document}